\documentclass[a4paper]{jpconf}
\usepackage{graphicx}
\begin{document}
\title{The unitary-model-operator approach to nuclear many-body problems}

\author{S Fujii$^{1,}$\footnote[3]{Present address, Center for Nuclear Study
(CNS), University of Tokyo, Wako Campus of RIKEN, Wako 351-0198, Japan},
R Okamoto$^{2}$, K Suzuki$^{2}$}

\address{$^{1}$ Department of Physics, University of Tokyo, Tokyo 113-0033,
Japan}
\address{$^{2}$ Department of Physics, Kyushu Institute of Technology,
Kitakyushu 804-8550, Japan}

\ead{sfujii@cns.s.u-tokyo.ac.jp}

\begin{abstract}
Microscopic nuclear structure calculations have been performed within
the framework of the unitary-model-operator approach. Ground-state and
single-particle energies are
calculated for nuclei around $^{14}$C, $^{16}$O and $^{40}$Ca with modern
nucleon-nucleon interactions.
\end{abstract}.

\section{Introduction}
Recently, {\it ab initio} nuclear structure calculations
starting with a realistic nucleon-nucleon (NN) interaction have been possible
beyond few-nucleon systems.
In this kind of study, the microscopic derivation of an effective interaction
from the realistic NN interaction allows one to perform the structure
calculation in a restricted model space.
One of the successful methods of the structure calculation in this direction
would be the no-core shell model (NCSM)~\cite{Navratil00,Navratil02}.
In the NCSM, the microscopic effective interaction (Hamiltonian) that takes
account of the short-range correlation of the original NN interaction
can be derived through a unitary transformation of the original Hamiltonian.
Although the shell-model diagonalization is done in a large model space
so that the final results do not depend on the model-space size,
such a model space is still much smaller than the huge Hilbert space
of the original Hamiltonian.
The microscopic understanding of nuclear structure has been growing through
the NCSM as well as the Green's function Monte Carlo
(GFMC)~\cite{Pieper01,Pieper05} in which
the original NN interaction and a three-nucleon force can be directly used.
However, although the GFMC and the NCSM are powerful method to describe
nuclear structure, the application of these methods to the nuclear structure
calculation may be limited to light nuclei up to $A\simeq 12$ due to
the present computer power.

If one wishes to describe heavier nuclei, one needs to have another method.
One of the promising methods may be the unitary-model-operator approach
(UMOA)~\cite{Suzuki94,Fujii00,Fujii04}.
The UMOA can be regarded as one of the coupled-cluster methods
(CCM)~\cite{Coester58,Coester60,Kuemmel78,Bartlett89,Mihaila00}
of Hermitian type. As for recent developments of the CCM in nuclear
theory, one may refer to Refs.~\cite{Kowalski04,Wloch05}.
In the UMOA, an energy independent and Hermitian effective interaction
is derived through a unitary transformation of the original
Hamiltonian~\cite{Okubo54,Suzuki82}
which is essentially the same as the unitary transformation used in the NCSM.
By doing the unitary transformation in a two-step procedure,
the structure calculation can be performed beyond $p$-shell nuclei.
So far, we have performed structure calculations for not only stable
nuclei around $^{16}$O~\cite{Fujii04} but also neutron-rich
oxygen isotopes~\cite{Fujii04c} and $\Lambda$ hypernuclei~\cite{Fujii00}.
In the following sections, we shall outline the calculation method of the
UMOA and present recent results for nuclei around $^{14}$C, $^{16}$O and
$^{40}$Ca.

\section{Method of calculation}

In the UMOA, the Hamiltonian $\tilde{H}$ to be considered is obtained
through a unitary transformation of the original Hamiltonian $H$ as
$\tilde{H}=e^{-S} He^{S}$.
The exponent $S$ is an anti-Hermitian operator and written as
$S={\rm arctanh}(\omega -\omega^{\dagger})$~\cite{Suzuki94,Suzuki82}
with a mapping operator $\omega=Q\omega P$ under the restrictive conditions
$PSP=QSQ=0$, where $P$ and $Q$ are the usual projection operators
and have the properties as $P+Q=1$, $P^{2}=P$, $Q^{2}=Q$ and $PQ=QP=0$.
Thus, the operator $\omega$ satisfies the relation
$\omega ^{2}={\omega ^{\dagger}}^{2}=0$. We should note here that
the unitary-transformation operator $U=e^{S}$ is also given by a block
form concerning the $P$ and $Q$ spaces as
\begin{equation}
\label{eq:U_block}
U=\left(
  \begin{array}{cc}
       P(1+\omega ^{\dagger}\omega)^{-1/2}P
    & -P\omega ^{\dagger}(1+\omega \omega ^{\dagger})^{-1/2}Q \\
       Q\omega (1+\omega ^{\dagger}\omega )^{-1/2}P
    &  Q(1+\omega \omega ^{\dagger})^{-1/2}Q
  \end{array}
\right)
\end{equation}
which agrees with the unitary transformation by
$\bar{\rm O}$kubo~\cite{Okubo54}.
By applying the above unitary transformation to a two-body subsystem
of the original Hamiltonian,
the two-body effective interaction $\tilde{v}_{12}$ of Hermitian type
is given by
\begin{equation}
\label{eq:eff_int}
\tilde{v}_{12}=U^{-1}(h_{0}+v_{12})U-h_{0},
\end{equation}
where $h_{0}$ is the one-body part and
$v_{12}$ is the bare two-body interaction.
The effective interaction in the $P$ and, if necessary, $Q$ spaces
is determined by solving the decoupling equation
$Q\tilde{v}_{12}P=P\tilde{v}_{12}Q=0$.
The actual method of calculating the matrix elements of $U$ and
$\tilde{v}_{12}$ with the harmonic-oscillator (h.o.) basis states
in the neutron-proton ($np$) formalism may be found in Ref.~\cite{Fujii04}.

\subsection{Two-step calculation for the effective interaction}

\begin{figure}[t]
\includegraphics[width=18pc]{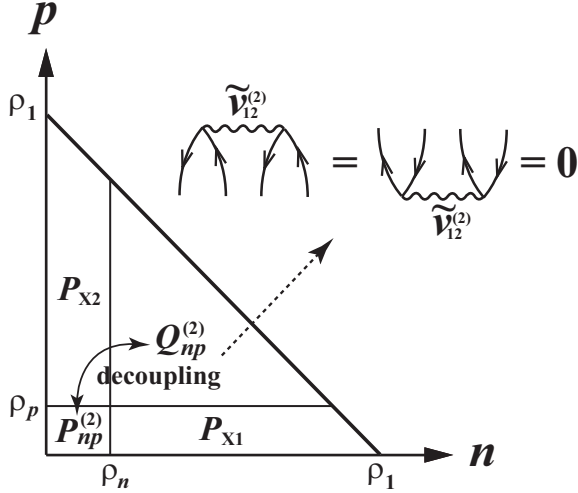}\hspace{2pc}%
\begin{minipage}[b]{18pc}\caption{\label{mspace2}The spaces consisting of
two-body states for the $np$ channel considered in the
second-step calculation for the effective interaction.}
\end{minipage}
\end{figure}

In order to make the structure calculation in an inexpensive way,
we perform the unitary transformation twice as follows.
First, we derive the two-body effective interaction in a large model space
to take into account the short-range correlation of the
original NN interaction.
The large model space consisting of two-body states is
specified by a boundary number $\rho _{1}$ which is given with the sets
of h.o. quantum numbers $\{ n_{a},l_{a}\}$ and $\{ n_{b},l_{b}\}$ of
the two-body states by $\rho _{1}=2n_{a}+l_{a}+2n_{b}+l_{b}$.
The value of $\rho _{1}$ is taken as large as possible so that the calculated
results do not depend on this value.
If we diagonalize the transformed Hamiltonian with the many-body
shell-model basis states
using the effective interaction in the large model space, that leads to the
NCSM. However, the calculations for heavier nuclei such as $^{16}$O
and $^{40}$O may not be practical in this manner
because the present computer cannot diagonalize
the huge matrix elements of the Hamiltonian
which is large enough to obtain the converged results.
If we intend to obtain only the energies of the ground states of closed-shell
nuclei and the single-particle (-hole) states in its neighboring nuclei,
it would be convenient to perform the unitary-transformation again
as follows.
We define a small {\it model} space $P_{np}^{(2)}$ and
its {\it complement} $Q_{np}^{(2)}$ by
separating the large model space in the previous procedure as shown in
Fig.~\ref{mspace2}.
The symbols $\rho _{n}$ and $\rho _{p}$ in Fig.~\ref{mspace2} stand for the
uppermost occupied states of the neutron and proton, respectively.
We here only show the $np$ channel.
It should be noted that the $P_{np}^{(2)}$ and
$Q_{np}^{(2)}$ spaces are considered on an equal footing
in this second-step calculation
when we solve the decoupling equation for the effective interaction
$\tilde{v}_{12}^{(2)}$ in this step as
$Q_{np}^{(2)}\tilde{v}_{12}^{(2)}P_{np}^{(2)}=0$
using the effective interaction determined in the first-step calculation
as an input.
Namely, we derive the effective interaction again in the large
model space.
However, by taking the $model$ space $P_{np}^{(2)}$ and its $complement$
$Q_{np}^{(2)}$ as shown in Fig.~\ref{mspace2},
the resultant effective interaction $\tilde{v}_{12}^{(2)}$ in this
second step has no vertices which induce two-particle two-hole
($2p2h$) excitation.
This is analogous to the Hartree-Fock (HF) condition which means
that an original Hamiltonian is transformed
so that the matrix elements for $1p1h$ excitation reduce to zero.
Although the vertices of the one-body non-diagonal matrix elements remain
in determining the effective interaction in the UMOA,
these non-diagonal matrix elements are diagonalized
at the end of the structure calculation.
We notice that the two-body effective interaction in the first- and
second-step calculations
is determined self-consistently with the one-body potential for both the
particle and hole states.
Although this procedure is not necessarily needed in determining the effective
interaction, such an effective interaction may be optimized for a restricted
model space so as to obtain a good unperturbed energy.

We here show an example of the effect of renormalizing $2p2h$ excitation
in the second-step calculation for the CD-Bonn potential~\cite{Machleidt96}.
The unperturbed ground-state energies of $^{16}$O for $\rho _{1}=12$
and $\hbar \Omega =15$ MeV using the effective interactions determined
in the first- and second-step calculations are
$-38.40$ and $-104.12$ MeV, respectively.
This means that a large amount of the $2p2h$ effect is renormalized
into the effective interaction in the small $model$ space
in the second-step calculation.
In order to obtain the final result including the rest correlation effects,
we diagonalize the transformed Hamiltonian with the shell-model basis states,
taking into account $1p1h$ excitation from the unperturbed ground state.
As for the closed-shell nucleus plus one-particle (one-hole) system,
the shell-model basis states are composed of the $1p$ and $2p1h$ states
($1h$ and $1p2h$ states). 
These correlation energies are added to the unperturbed ground-state energies,
and then we obtain the final results of the ground-state energies
of the closed-shell nucleus and the closed-shell nucleus plus one-particle
(one-hole) system.
We note here that there remain some correction terms to be evaluated,
such as the three-body cluster terms.
Although the three-body cluster effect is essentially small,
that may have a significant effect for some particular cases.
Actually, we have found that the evaluation of the three-body
cluster effect plays an important role to describe shell structure
in neutron-rich oxygen isotopes~\cite{Fujii04c}.
However, generally speaking, the magnitude of the three-body cluster
effect is much smaller than the two-body cluster effect, and thus
the cluster expansion of the transformed Hamiltonian is justified from
the viewpoint of the perturbative expansion. In the next section,
we show some of the recent results which do not include the three-body
cluster effect. The results including the three-body cluster effect
will be reported elsewhere in the near future.

\section{Results and discussion}

\begin{table}[t]
\caption{\label{tab:O16_gs} The calculated ground-state
energies with the $1p1h$ effect $E_{\rm g.s.}$
and the binding energies per nucleon $BE/A$
of $^{16}$O.
All energies are in MeV.}
\begin{center}
    \begin{tabular}{cccccc}
\br
      $^{16}$O        &   Nijm 93  &    Nijm I   &    N$^{3}$LO   &   CD Bonn   &    Expt.   \\ \mr
      $ E_{\rm g.s.} $ & $ -99.69 $ & $ -104.25 $ & $ -110.00 $ & $ -115.61 $ & $ -127.62 $ \\
      $ BE/A $        & $   6.23 $ & $    6.52 $ & $    6.88 $ & $    7.23 $ & $    7.98 $ \\
\br
    \end{tabular}
\end{center}
\end{table}

\begin{figure}[b]
\begin{center}
\includegraphics[width=150mm]{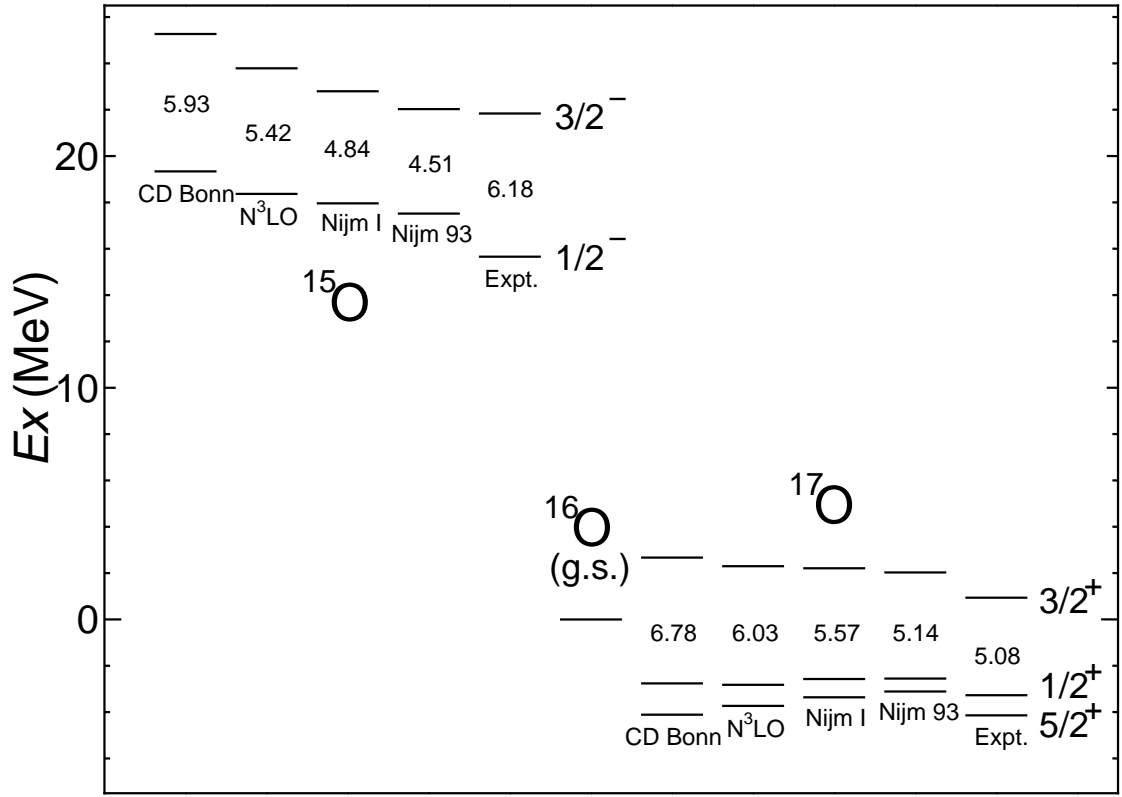}
\end{center}
\caption{\label{oxygen}The calculated single-particle and single-hole energy
levels in $^{17}$O and $^{15}$O, respectively, with modern NN interactions
relative to the ground state of $^{16}$O.}
\end{figure}

First of all, we show calculated ground-state energies of $^{16}$O
with the $1p1h$ effect with modern NN interactions in Table~\ref{tab:O16_gs}.
The final results for the Nijm-93,
Nijm-I~\cite{Stoks94}, the CD-Bonn~\cite{Machleidt96} and
the chiral N$^{3}$LO~\cite{Entem03} potentials are tabulated
together with the experimental values.
The binding energies per nucleon are also shown.
The Coulomb force is used commonly.
The results for the Nijm~93 and the CD~Bonn are the least and most attractive,
respectively, of the four potentials.
This tendency can also be observed in the Faddeev-Yakubovsky calculations
for $^{4}$He by Nogga {\it et al}~\cite{Nogga02}.
It is seen that the calculated ground-state energies are less bound
than the experimental value.
In the present results, higher-order correlations
such as the three-body cluster terms have not been evaluated.
If we include the effect of the three-body cluster terms,
the calculated results become more attractive by a few MeV.
For example, in the case of the CD-Bonn potential,
the three-body cluster effect is about $-4$ MeV.
However, even if we add this value to the ground-state energy
in Table~\ref{tab:O16_gs},
the result is less attractive than the experimental value.
In the present calculation, a genuine three-body force is not included.
The inclusion of the real three-body force and the higher-order
many-body correlations would compensate for the discrepancies
between the experimental and calculated values.
Such a study remains as an important task for a deeper understanding of
nuclear ground-state properties in the UMOA.

In Fig.~\ref{oxygen}, calculated single-particle levels in $^{17}$O
and single-hole levels in $^{15}$O relative to the ground state of
$^{16}$O are illustrated.
The values in Fig.~\ref{oxygen} are the spin-orbit splitting energies
for hole states in $^{15}$O and particle states in $^{17}$O.
We see that the calculated spin-orbit splittings for the hole states
in $^{15}$O are smaller than
the experimental value.
As for the spin-orbit splittings for the particle states $^{17}$O,
as opposed to the hole-state case, the energies are larger than the
experimental value. This may be due to an insufficient treatment
for the $3/2^{+}$ unbound state in the present calculation.
We have used only the h.o. states as the basis states.
However, we may say that the calculated spin-orbit splittings for the hole and
particle states in nuclei around $^{16}$O are, on the whole, not very
different from the experimental values though the results somewhat
depend on the interactions employed.
The magnitudes of remaining discrepancies may be reduced
if we include the genuine three-body force in the calculation
and evaluate the higher-order cluster terms.

\begin{figure}[t]
\begin{minipage}{18pc}
\includegraphics[width=18pc]{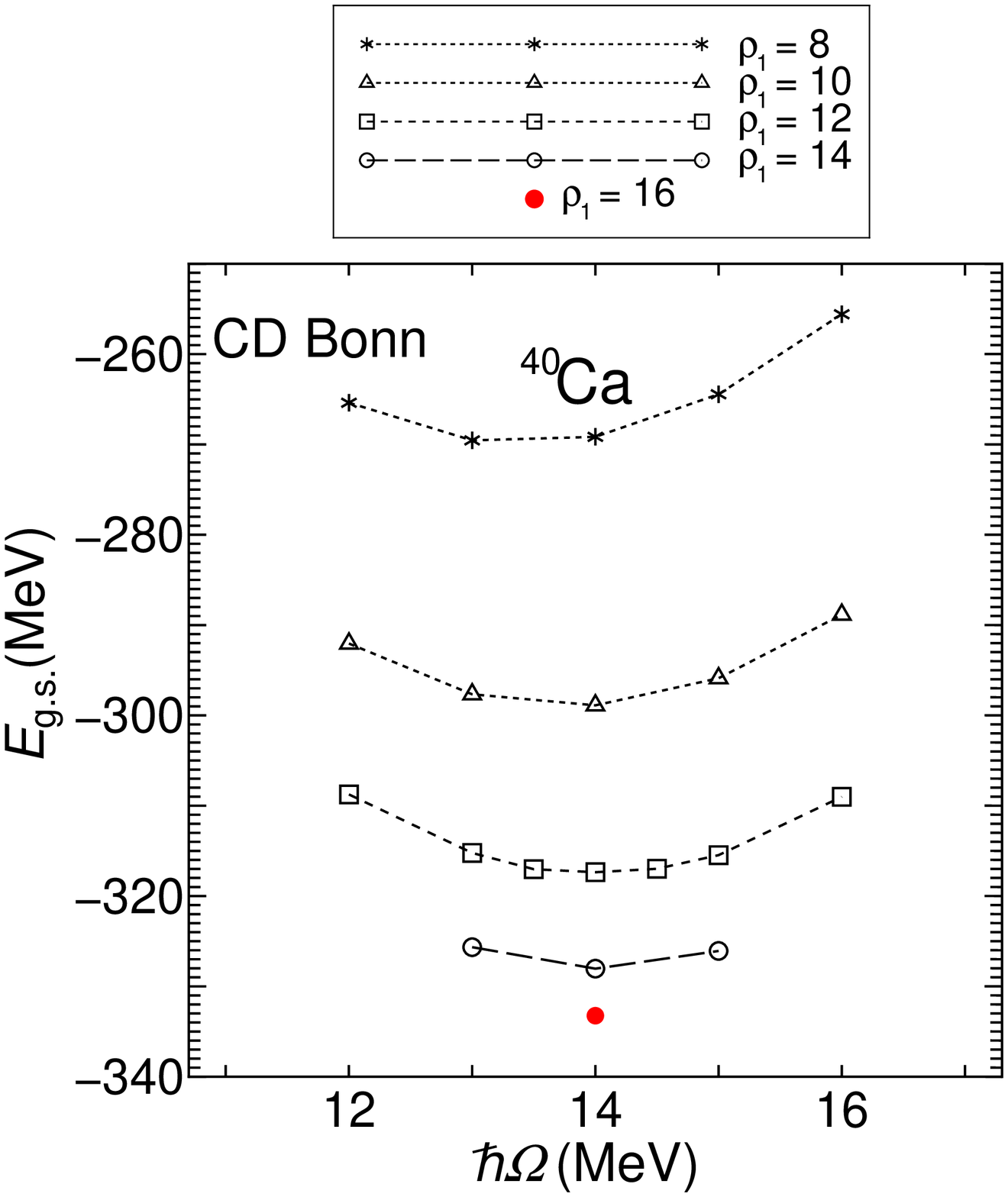}
\caption{\label{Ca40}The $\rho _{1}$ and $\hbar \Omega$ dependences of
the calculated ground-state energies of $^{40}$Ca for the CD-Bonn potential.}
\end{minipage}\hspace{2pc}%
\begin{minipage}{18pc}
\includegraphics[width=18pc]{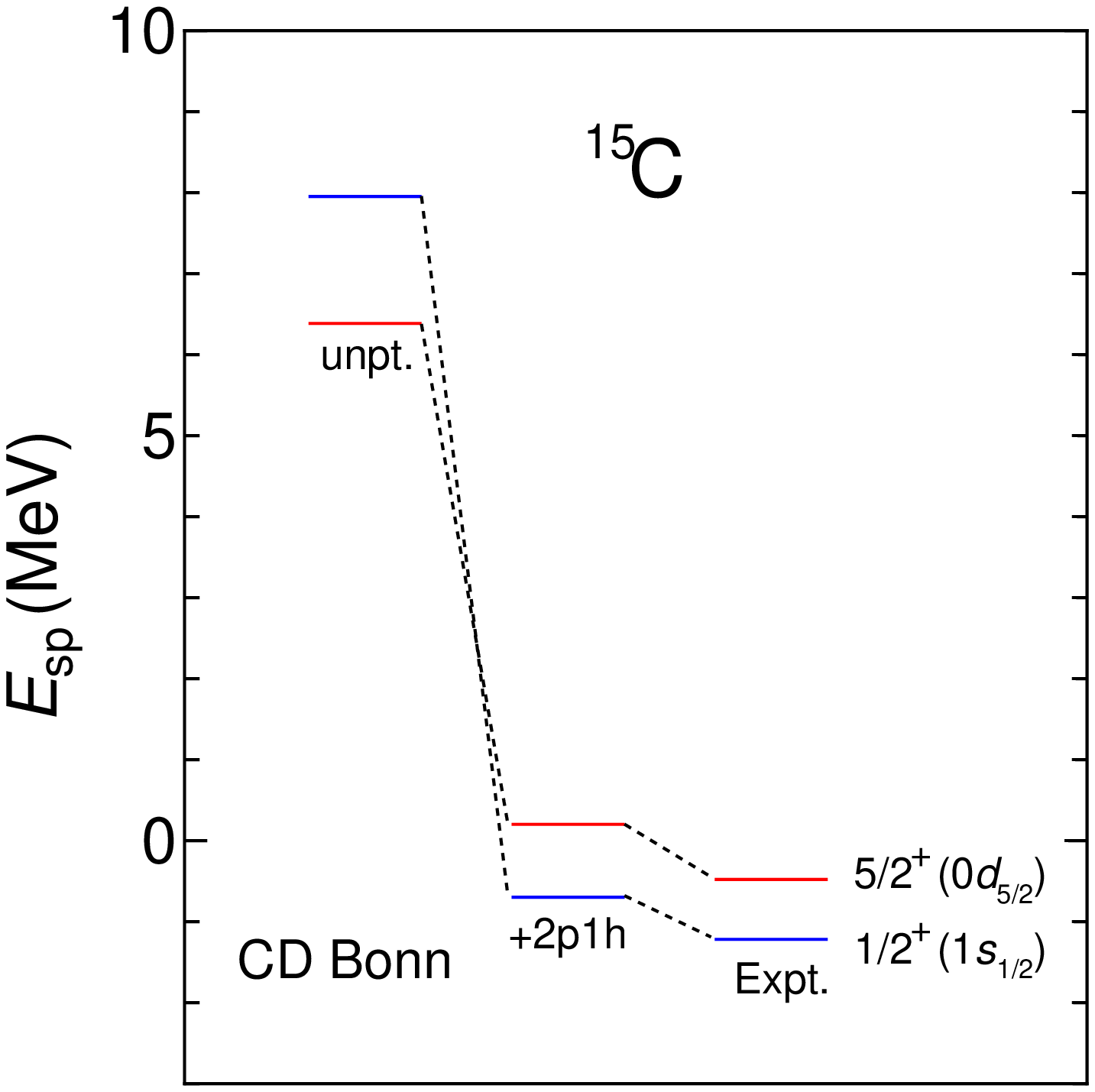}
\caption{\label{C15}The calculated single-particle energies in $^{15}$C
for the CD-Bonn potential.}
\end{minipage} 
\end{figure}

We are now trying to do calculations for heavier systems.
Here we show some of the preliminary results for $^{40}$Ca.
Figure~\ref{Ca40} shows the $\rho _{1}$ and $\hbar \Omega$ dependences
of the ground-state energy with the $1p1h$ effect for the CD-Bonn potential.
It is seen that a fairly convergent result is obtained at $\rho _{1}=16$
and $\hbar \Omega =14$ MeV. Though we are now calculating for $\rho _{1}=18$,
the difference of the results between $\rho _{1}=16$ and $\rho _{1}=18$
would be a few MeV at most.
In the case of $^{16}$O, the convergent result
can be obtained at $\rho _{1}=14$.
Since we are now calculating heavier systems, we need a larger model space
to obtain the convergent result.

In Fig.~\ref{C15}, we show calculated single-particle energies
in a neutron-rich nucleus $^{15}$C together with the experimental
values.
The result for ``unpt." denotes the unperturbed single-particle energy
which is defined as the h.o. kinetic energy plus the self-consistent
one-body potential determined in the second-step calculation.
In $^{15}$C, the ordering of the experimental single-particle $1/2^{+}$ and
$5/2^{+}$ states are opposite to the case of $^{17}$O.
However, our result for ``unpt." does not reproduce this tendency,
and the single-particle levels are rather repulsive to the experimental
values.
However, if we see the results with the $2p1h$ effect by the diagonalization
as shown in the middle of Fig.~\ref{C15},
the results become more attractive and the two levels are reversed,
and then a good agreement with the experimental value is obtained.
We should remark, however, that the three-body cluster terms remain to
be evaluated.
As has been reported for neutron-rich oxygen isotopes in Ref.~\cite{Fujii04c},
the three-body cluster has a non-negligible effect on particle states of
loosely-bound neutron-rich systems.
Therefore, for the complete study, we have to evaluate the three-body cluster
terms and also investigate the effect of the genuine three-body force
though the present two-body cluster approximation should be a good treatment.

As has been shown before,
the UMOA is a useful many-body theory to microscopically
describe nuclear structure near closed-shell nuclei.
The ground-state energy and the single-particle (-hole) energy
can be calculated systematically for not only $N\simeq Z$ nuclei
but also neutron-proton asymmetric systems beyond $p$-shell nuclei.
Forthcoming new facilities for accelerating RI beams in the world
will reveal new and exciting phenomena such as the
change of shell structure in asymmetric nuclei.
The UMOA has a possibility to develop new structures
from a microscopic point of view in investigating the
exotic nuclear systems.

\ack

This work was supported by a Grant-in-Aid for Scientific Research (C)
(Grant No. 15540280) from Japan Society for the Promotion of Science
and a Grant-in-Aid for Specially Promoted Research (Grant No. 13002001)
from the Ministry of Education, Culture, Sports, Science and Technology
in Japan.

\end{document}